\documentclass[sigconf]{acmart}
\usepackage[utf8]{inputenc}
\usepackage{subcaption}
\usepackage{wrapfig}
\usepackage{makecell}
\usepackage{verbatim}
\usepackage{xcolor}
\usepackage{url}
\usepackage{listings}
\usepackage{setspace}
\usepackage{algorithm}
\usepackage{enumitem}
\usepackage[noend]{algpseudocode}
\usepackage{float}

\usepackage{amsthm}

\algblock{Input}{EndInput}
\algblock{Output}{EndOutput}

\usepackage{xparse}
\usepackage{mathtools}
\usepackage{multirow}
\usepackage{datapie}
\usepackage{adjustbox}
\usepackage[most]{tcolorbox}
\usepackage{pgf}
\usepackage{eqparbox}
\newdimen{\algindent}
\algnewcommand\LeftComment[2]{%
\hspace{#1\algindent}$\triangleright$ #2 \hfill %
}

\newcommand{\name}{\textit{PALM}}
\newcommand{\hammer}{CoqHammer}

\newcommand{\gpt}{GPT-3.5}
\newcommand{\gptfour}{GPT-4o}
\newcommand{\llamaseven}{Llama-3-70b-Instruct}
\newcommand{\llamaeight}{Llama-3-8b-Instruct}
\newcommand{\namegpt}{\name{} (GPT-3.5)}

\newcommand{\passport}{\textit{Passport}}
\newcommand{\proverbot}{\textit{Proverbot9001}}

\newcommand{\dspfull}{\textit{Draft, Sketch, and Prove (DSP)}}
\newcommand{\dsp}{\textit{DSP}}

\newcommand{\varnoref}{\name{}\_$_{ref}$}
\newcommand{\varnoname}{\name{}\_$_{rename}$}
\newcommand{\varnobullet}{\name{}\_$_{bullet}$}
\newcommand{\varnoaug}{\name{}\_$_{aug}$}

\newcommand{\varnoretriever}{\name{}\_$_{retriever}$}
\newcommand{\varnobacktrack}{\name{}\_$_{backtrack}$}

\newcommand{\finding}[1]{\begin{tcolorbox}[enhanced, frame hidden, boxsep=0pt]\emph{#1}\end{tcolorbox}}

\newcommand{\typeOne}{Invalid reference}
\newcommand{\typeTwo}{Redundant introductions}
\newcommand{\typeThree}{Bullet misuse}
\newcommand{\typeFour}{Wrong theorem application}
\newcommand{\typeFive}{Incorrect rewrite}
\newcommand{\typeSix}{Tactic misuse}

\newcommand{\repairOne}{Reference replacement}
\newcommand{\repairTwo}{Renaming}
\newcommand{\repairThree}{Bullet transformation}
\newcommand{\repairFour}{Premise augmentation}

\usepackage{tikz}
\usetikzlibrary{shapes.geometric, arrows, calc}
\tikzset{
    proofstate/.style={
        draw,
        rectangle,
        fill=gray!20,
        text width=4.5cm,
        align=center,
        minimum height=2cm,
        inner sep=5pt
    },
    proofline/.style={
        draw,
        thick
    }
}

\AtBeginDocument{%
  }

\copyrightyear{2024}
\acmYear{2024}
\setcopyright{rightsretained}
\acmConference[ASE '24]{39th IEEE/ACM International Conference on Automated Software Engineering }{October 27-November 1, 2024}{Sacramento, CA, USA} \acmBooktitle{39th IEEE/ACM International Conference on Automated Software Engineering (ASE '24), October 27-November 1, 2024, Sacramento, CA, USA} \acmDOI{10.1145/3691620.3695521}
\acmISBN{979-8-4007-1248-7/24/10}

\begin{document}

\newcommand{\calnameprecapp}[4]{
    \newcommand{#1}{\pgfmathsetmacro{\result}{#2}\pgfmathprintnumber[fixed zerofill, precision=#3]{\result}#4}
}
\def\totalnum{10842}

\def\succnum{4377}
\def\succnumfour{4614}
\def\succnumourstwo{5154}
\def\succnumllama{4155}
\def\succnummistral{3433}
\def\succnumpassport{1561}
\def\succnumproverbot{1849}
\def\succnumcoqhammer{3182}
\def\succnumdsp{2478}
\def\succnumcombined{3616}

\def\succnuminitial{402}
\calnameprecapp{\succrateinitial}{\succnuminitial/\totalnum*100}{1}{\%}

\def\succnuminitialfour{689}
\calnameprecapp{\succrateinitialfour}{\succnuminitialfour/\totalnum*100}{1}{\%}

\def\succnuminitialllamaseven{386}
\calnameprecapp{\succrateinitialllamaseven}{\succnuminitialllamaseven/\totalnum*100}{1}{\%}

\def\succnuminitialllamaeight{7}
\calnameprecapp{\succrateinitialllamaeight}{\succnuminitialllamaeight/\totalnum*100}{1}{\%}

\calnameprecapp{\succrate}{\succnum/\totalnum*100}{1}{\%}
\calnameprecapp{\succratefour}{\succnumfour/\totalnum*100}{1}{\%}
\calnameprecapp{\succratellama}{\succnumllama/\totalnum*100}{1}{\%}
\calnameprecapp{\succratemistral}{\succnummistral/\totalnum*100}{1}{\%}
\calnameprecapp{\succrateourstwo}{\succnumourstwo/\totalnum*100}{1}{\%}

\calnameprecapp{\succratepassport}{\succnumpassport/\totalnum*100}{1}{\%}
\calnameprecapp{\succrateproverbot}{\succnumproverbot/\totalnum*100}{1}{\%}
\calnameprecapp{\succratecoqhammer}{\succnumcoqhammer/\totalnum*100}{1}{\%}
\calnameprecapp{\succratedsp}{\succnumdsp/\totalnum*100}{1}{\%}
\calnameprecapp{\succratecombined}{\succnumcombined/\totalnum*100}{1}{\%}

\calnameprecapp{\outpassport}{(\succnum-\succnumpassport)/\succnumpassport*100}{1}{$\%$}
\calnameprecapp{\outproverbot}{(\succnum-\succnumproverbot)/\succnumproverbot*100}{1}{$\%$}
\calnameprecapp{\outcoqhammer}{(\succnum-\succnumcoqhammer)/\succnumcoqhammer*100}{1}{\%}
\calnameprecapp{\outdsp}{(\succnum-\succnumdsp)/\succnumdsp*100}{1}{\%}
\calnameprecapp{\outcombined}{(\succnum-\succnumcombined)/\succnumcombined*100}{1}{\%}

\calnameprecapp{\outpassportllama}{(\succnumllama-\succnumpassport)/\succnumpassport*100}{1}{$\%$}
\calnameprecapp{\outproverbotllama}{(\succnumllama-\succnumproverbot)/\succnumproverbot*100}{1}{$\%$}
\calnameprecapp{\outcoqhammerllama}{(\succnumllama-\succnumcoqhammer)/\succnumcoqhammer*100}{1}{\%}
\calnameprecapp{\outcombinedllama}{(\succnumllama-\succnumcombined)/\succnumcombined*100}{1}{\%}

\def\succnumnoref{4249}
\def\succnumnoname{4175}
\def\succnumnobullet{4225}
\def\succnumnoaug{4094}
\def\succnumnohammer{702}
\def\succnumnobacktrack{702}
\def\succnumnoretriever{4147}

\calnameprecapp{\succratenoref}{\succnumnoref/\totalnum*100}{1}{\%}
\calnameprecapp{\succratenoname}{\succnumnoname/\totalnum*100}{1}{\%}
\calnameprecapp{\succratenobullet}{\succnumnobullet/\totalnum*100}{1}{\%}
\calnameprecapp{\succratenoaug}{\succnumnoaug/\totalnum*100}{1}{\%}
\calnameprecapp{\succratenohammer}{\succnumnohammer/\totalnum*100}{1}{\%}
\calnameprecapp{\succratenobacktrack}{\succnumnobacktrack/\totalnum*100}{1}{\%}
\calnameprecapp{\succratenoretriever}{\succnumnoretriever/\totalnum*100}{1}{\%}

\calnameprecapp{\outnoref}{(\succnum-\succnumnoref)/\succnumnoref*100}{1}{\%}
\calnameprecapp{\outnoname}{(\succnum-\succnumnoname)/\succnumnoname*100}{1}{\%}
\calnameprecapp{\outnobullet}{(\succnum-\succnumnobullet)/\succnumnobullet*100}{1}{\%}
\calnameprecapp{\outnoaug}{(\succnum-\succnumnoaug)/\succnumnoaug*100}{1}{\%}
\calnameprecapp{\outnohammer}{(\succnum-\succnumnohammer)/\succnumnohammer}{1}{$\times$}
\calnameprecapp{\outnobacktrack}{(\succnum-\succnumnobacktrack)/\succnumnobacktrack*100}{1}{\%}
\calnameprecapp{\outnoretriever}{(\succnum-\succnumnoretriever)/\succnumnoretriever*100}{1}{\%}

\title{Proof Automation with Large Language Models}
% Neuro-Symbolic Proof Automation with Large Language Model
% Generating Formal Proofs using Large Language Models and Repair via Symbolic Methods
% Generation with Large Language Models and Repair Using Symbolic Methods
% Enhancing Proof Automation: Generation with Large Language Models and Repair via Symbolic Methods
\definecolor{lightgray}{gray}{0.97}
\definecolor{keywordcolor}{rgb}{0.13, 0.13, 1} % Blue
\definecolor{keywordcolor2}{rgb}{0.5,0,0.5}

\lstdefinelanguage{Coq}{
  morekeywords={Type, Prop, Definition, Lemma, Theorem, Proof, Qed, Fixpoint},
  % morekeywords={induction, with, shelve, apply, fun, let, in, forall, exists, match},
  sensitive=true, % keywords are case-sensitive
}

\lstdefinestyle{coqline}{
    breakatwhitespace=false,
    breaklines=true,
    captionpos=t,
    keepspaces=true,
    basicstyle=\ttfamily\footnotesize,
    showspaces=false,
    showstringspaces=false,
    showtabs=false,
    tabsize=2,
    language=Coq,
    keywordstyle=\color{keywordcolor},
    breaklines=true,
    frame=tb,
    framesep=1pt,
    framerule=0.5pt,
    rulecolor=\color{black},
    aboveskip=0pt,
    belowskip=0pt,
    escapeinside={(*@}{@*)},
    numbersep=0pt,
    numbers=left,
    numberstyle=\tiny,
  }

\lstdefinestyle{coq}{
    breakatwhitespace=false,
    breaklines=true,
    captionpos=t,
    keepspaces=true,
    basicstyle=\ttfamily\footnotesize,
    showspaces=false,
    showstringspaces=false,
    showtabs=false,
    tabsize=2,
    language=Coq,
    keywordstyle=\color{keywordcolor},
    breaklines=true,
    frame=single,
    framesep=1pt,
    framerule=0.5pt,
    rulecolor=\color{black},
    % aboveskip=0pt,
    belowskip=-2pt,
    escapeinside={(*@}{@*)},
  }

\author{Minghai Lu}
\email{lu1074@purdue.edu}
\orcid{0009-0001-0136-3204}
\affiliation{%
  \institution{Purdue University}
  \city{West Lafayette}
  \state{IN}
  \country{USA}
}

\author{Benjamin Delaware}
\email{bendy@purdue.edu}
\orcid{0000-0002-1016-6261}
\affiliation{%
  \institution{Purdue University}
  \city{West Lafayette}
  \state{IN}
  \country{USA}
}

\author{Tianyi Zhang}
\email{tianyi@purdue.edu}
\orcid{0000-0002-5468-9347}
\affiliation{%
  \institution{Purdue University}
  \city{West Lafayette}
  \state{IN}
  \country{USA}
}

%%
%% Article type: Research, Review, Discussion, Invited or position
% \acmArticleType{Research}
%%
%% Links to code and data
\acmCodeLink{https://github.com/borisveytsman/acmart}
\acmDataLink{htps://zenodo.org/link}


\begin{abstract}
  Interactive theorem provers such as Coq are powerful tools to formally
  guarantee the correctness of software. However, using these tools
  requires significant manual effort and expertise.
% Existing proof automation techniques either cannot perform specific proof steps like inductive reasoning, or require training on substantial domain-specific data.
  While Large Language Models (LLMs) have shown promise in automatically
  generating informal proofs in natural language, they
  are less effective at generating formal proofs in interactive
  theorem provers.
  % Even the most advanced LLMs can only prove a few theorems due to
  % various errors.
  In this paper, we conduct a formative study to identify common
  mistakes made by LLMs when asked to generate formal proofs. By
  analyzing 520 proof generation errors made by \gpt{}, we found that \gpt{} often identified the correct high-level structure of a proof, but struggled to get the lower-level details
  correct. Based on this insight, we propose \name{}, a novel
  generate-then-repair approach that first prompts
  an LLM to generate an initial proof and then leverages targeted
  symbolic methods to iteratively repair low-level problems. We
  evaluate \name{} on a large dataset that includes more than 10K
  theorems. Our results show that \name{} significantly outperforms
  other state-of-the-art approaches, successfully proving \outdsp{} to
  \outpassport{} more theorems. Moreover, \name{} proves 1270 theorems
  beyond the reach of existing approaches. We also demonstrate the
  generalizability of \name{} across different LLMs.

% Moreover, \name{} integrates the components with a flexible search framework which can be expanded
% In particular, \name{} adopts a premise retriever to select relevant premises such as lemmas and definitions, to enhance the performance of LLMs. Moreover, it employs a set of repair mechanisms to correct errors made by LLMs. All these components are integrated within a flexible framework, that can be easily adapted to other ITPs and expanded with additional repair mechanisms.
% , offering flexibility and generalizability.
\end{abstract}

\maketitle

\section{Introduction}
Correctness is crucial to software systems. Interactive theorem
provers (ITPs) such as Coq~\cite{coq}, Isabelle~\cite{isabelle} and
Lean~\cite{lean}, are powerful tools for providing semantically rich
guarantees about software. In an ITP, users can state and prove formal theorems about a program; these proofs are then mechanically checked by the ITP, providing a strong, foundational guarantee about its correctness. This strategy has been successfully applied to several application domains, including compilers~\cite{CompCert-ERTS-2018}, distributed systems~\cite{verdi}, and OS kernels~\cite{klein2009sel4}. While powerful, this approach comes at a cost, as users must supply a \emph{proof script} that helps the ITP construct the proof of the desired theorem. Constructing these proof scripts can require considerable effort. For example, it took 6 person-years to write 100,000 lines of Coq proof scripts to verify the CompCert C compiler~\cite{CompCert-ERTS-2018}.

Many proof automation techniques have been proposed to reduce the
effort required by ITPs. These techniques mainly fall into two
categories: symbolic methods~\cite{coqhammer,
  sledgehammer, kaliszyk2015hol, coqauto} and machine
learning methods~\cite{yang2019learning, first2020tactok,
  first2022diversity, sanchez2023passport}. Symbolic methods use a
combination of previously established theorems and external automated
theorem provers (ATPs), such as Z3~\cite{z3} and CVC5~\cite{cvc5}, to
automate the proof of a theorem. While effective, these approaches are constrained by their inability to perform higher-order and inductive reasoning, limiting their ability to prove complex theorems.
Machine learning methods utilize models to predict the next proof step in a heuristic-guided search process.
These methods do not have the same limitations as symbolic
approaches but require a significant amount of training data~\cite{yang2019learning,
  first2020tactok,
  first2022diversity}.

Recently, pretrained Large Language Models (LLMs) have shown promise
in generating informal natural language proofs~\cite{wu2022autoformalization}, suggesting a potential to further
improve existing proof automation approaches. Unfortunately, even state-of-the-art LLMs are ineffective at generating formal proofs in one shot: \gpt{} proves \succrateinitial{} of theorems in our evaluation, and \llamaseven{} proves \succrateinitialllamaseven{}.
In order to understand why, we have conducted a formative study to analyze mistakes that \gpt{} made when generating formal proofs. In this study, we analyzed 579 theorems of varied complexity and identified seven categories of errors. Overall, we found that while \gpt{} often produced proofs with the right high-level structure, it struggled getting lower-level details of these proofs correct. Promisingly, we also observed that many of these errors can be potentially fixed using symbolic methods, including heuristic-based search and proof repair.

Guided by this formative study, we propose \name{}, a novel generate-then-repair approach that combines LLMs and symbolic methods. Our key insight is to use LLMs to generate an initial proof that is likely to have the correct high-level structure, and then use targeted symbolic methods to iteratively repair low-level problems related to individual proof steps. \name{} relies on four repair mechanisms that target the common types of errors identified in our formative study. If our repair mechanisms fail, \name{} uses a backtracking procedure to regenerate previous proof steps in an attempt to fix errors in the high-level proof structure.
Although \name{} targets Coq, its underlying principles can be applied
to other ITPs, such as Isabelle~\cite{isabelle} and Lean~\cite{lean}.

To evaluate the effectiveness of our approach, we have conducted an
extensive evaluation using the CoqGym dataset~\cite{yang2019learning} with \totalnum{} theorems. Our results suggest that \name{} can successfully prove \succrate{} of the theorems, significantly outperforming the state-of-the-art methods \passport{}~\cite{sanchez2023passport},
\proverbot{}~\cite{sanchez2020generating} and
\dspfull{}~\cite{jiang2022draft}, which only prove
\succratepassport{}, \succrateproverbot{} and \succratedsp{} theorems,
respectively. Moreover, we have conducted experiments to demonstrate the effectiveness of each component in \name{} and the generalizability of \name{} across different LLMs.

In summary, this paper presents the following contributions:
\begin{enumerate} 
\item We conduct a formative study to identify the common errors made by GPT-3.5 while proving theorems in Coq. 
\item We propose \name{}, a novel proof automation approach that combines LLMs and symbolic methods in a generate-then-repair pipeline.
\item We evaluate \name{} on a large dataset and demonstrate that \name{} significantly outperforms existing methods. An artifact containing the source code of \name{} and a replication package is publicly available~\cite{oursgithub}.
\end{enumerate}

\section{Preliminaries}

\subsection{Interactive Theorem Proving in Coq}
The Coq proof assistant~\cite{coq} is a popular tool for developing
machine-checked proofs of mathematical theorems and verifying complex
software systems. Coq helps users interactively construct these proofs
using a set of proof \textit{tactics}.
This section first introduces the basic concepts of interactive proof
development in Coq, and then illustrates the process via an example
theorem shown in Figure~\ref{eg:theorem}.
\begin{figure}[t]
\centering
\lstset{style=coqline}
\begin{lstlisting}
 Theorem add_comm : forall n m : nat, n + m = m + n.
 Proof.
   intros n m.
   induction n.
   -
   auto.
   -
   simpl.
   rewrite IHn.
   apply plus_n_Sm.
 Qed.
\end{lstlisting}
% \vspace{-10pt}
\caption{\label{eg:theorem} A Coq theorem stating that natural number
  addition is commutative, and a proof of this statement.}
% \vspace{-10pt}
\end{figure}

\textbf{Theorems:} In Coq, the definition of a \textit{theorem}
typically starts with the keyword \texttt{Theorem} or \texttt{Lemma},
followed by its name and the theorem statement.
Figure~\ref{eg:theorem} shows the theorem \textit{add\_comm}, which
states that natural number addition is commutative.\footnote{The type of natural numbers in Coq is \texttt{nat}.} This is then followed by a \emph{proof script}, a sequence of tactics that explain how to build a proof of the desired statement. Proof scripts are typically developed in an interactive proof mode. Processing the first line of Figure~\ref{eg:theorem} causes Coq to enter proof mode. During the proof process, users can freely reuse previously proven theorems.

\textbf{Proof States:} In proof mode, Coq's interface displays the
current \textit{proof state}, i.e., a list of unproven $goals$. Each
of these goals is a pair of a local context $lc$ and an outstanding proof obligation $st$. A local context includes hypotheses and
assumptions that can be used to prove $st$; these are distinct from
the set of previously proven theorems, which are part of the global
context. Figure~\ref{eg:tree} shows the intermediate proof states that
appear during the proof of \textit{add\_comm}: each listing shows the
proof states shown to the user after processing each tactic in
Figure~\ref{eg:theorem}. Following the conventions of Coq's user
interface, the local context is shown above the double line, and the
current proof obligation is shown below.

\begin{figure}[t]
\centering
\begin{subfigure}[h]{0.45\textwidth}
\lstset{style=coq}
\begin{lstlisting}
================================
forall n m : nat, n + m = m + n
\end{lstlisting}
\caption{Proof state at the start.}
\label{eg:a}
\end{subfigure}

\vspace{5pt}
\begin{subfigure}[h]{0.45\textwidth}
\lstset{style=coq}
\begin{lstlisting}
n, m: nat
================================
n + m = m + n
\end{lstlisting}
\caption{Proof state after Figure~\ref{eg:theorem} Line 3 (\texttt{intros n m}).}
\label{eg:b}
\end{subfigure}

\vspace{5pt}
\begin{subfigure}[h]{0.45\textwidth}
\lstset{style=coq}
\begin{lstlisting}
m: nat
================================
(1/2)
0 + m = m + 0
(2/2)
S n + m = m + S n
\end{lstlisting}
\caption{Proof state after Figure~\ref{eg:theorem} Line 4 (\texttt{induction n}).}
\label{eg:c}
\end{subfigure}

\vspace{5pt}
\begin{subfigure}[h]{0.45\textwidth}
\lstset{style=coq}
\begin{lstlisting}
m: nat
================================
0 + m = m + 0
\end{lstlisting}
\caption{Proof state after Figure~\ref{eg:theorem} Line 5 (the first subgoal).}
\label{eg:d}
\end{subfigure}

\vspace{5pt}
\begin{subfigure}[h]{0.45\textwidth}
\lstset{style=coq}
\begin{lstlisting}
n, m: nat
IHn: n + m = m + n
================================
S n + m = m + S n
\end{lstlisting}
\caption{Proof state after Figure~\ref{eg:theorem} Line 7 (the second subgoal).}
\label{eg:e}
\end{subfigure}

\vspace{5pt}
\begin{subfigure}[h]{0.45\textwidth}
\lstset{style=coq}
\begin{lstlisting}
n, m: nat
IHn: n + m = m + n
================================
S (n + m) = m + S n
\end{lstlisting}
\caption{Proof state after Figure~\ref{eg:theorem} Line 8 (\texttt{simpl}).}
\label{eg:f}
\end{subfigure}

\vspace{5pt}
\begin{subfigure}[h]{0.45\textwidth}
\lstset{style=coq}
\begin{lstlisting}
n, m: nat
IHn: n + m = m + n
================================
S (m + n) = m + S n
\end{lstlisting}
\caption{Proof state after Figure~\ref{eg:theorem} Line 9 (\texttt{rewrite IHn}).}
\label{eg:g}
\end{subfigure}
\vspace{-10pt}
\caption{Proof state after the execution of each tactic in the proof of addition's commutativity.}
\label{eg:tree}
\vspace{-10pt}
\end{figure}

\textbf{Tactics:} Tactics specify strategies for decomposing the
current proof obligation into a set of simpler subgoals, in order to
eventually produce a complete proof. Conceptually, a tactic $t$ is a
state-transition function: $t \in S\times \sum \rightarrow S'$, where
$S$ is a goal, $\sum$ is a set of
arguments if any, and $S'$ is the set of resulting goals. As an
example, the tactic \texttt{induction n} on Line 4 of
Figure~\ref{eg:theorem} tells Coq to do induction on the natural
number \texttt{n} in the local context. Processing this tactic
transforms the proof state in Figure~\ref{eg:b} to the proof state in
Figure~\ref{eg:c}, which has two subgoals: (1) a base case in which
\texttt{n} is \texttt{0}, and (2) an inductive case in which
\texttt{n} is an arbitrary natural number.\footnote{The term \texttt{S n} is equivalent to \texttt{n + 1}.} Note that Coq only displays the local context of the first goal when there are multiple goals. Importantly, a tactic can fail if, for example, it is applied to a proof state of the wrong form or it is supplied with wrong arguments. Coq reports the failure back to the user when this occurs.

\textbf{Proofs:} A proof of a theorem consists of a sequence of
tactics that transform the initial goal, i.e., the theorem statement,
into subgoals until none remain. The beginning and end of a proof are delimited by the \texttt{Proof} and \texttt{Qed} commands (Lines 2 and 11 of Figure~\ref{eg:theorem}). The latter command prompts Coq's kernel to check that no outstanding proof obligations remain. If so, Coq exits proof mode with success and the theorem is added to the global context.

We now illustrate these concepts using the example of
\texttt{add\_comm} in Figure~\ref{eg:theorem}. At the beginning of the
proof (Line 2), the proof state consists of a single goal that
corresponds to the top-level theorem statement (Figure~\ref{eg:a}). At Line 3, the tactic \texttt{intros n m} tells Coq to move the universally quantified variables \texttt{n} and
\texttt{m} into the local context (Figure~\ref{eg:b}). Then, the aforementioned \texttt{induction n} tactic performs induction on
\texttt{n} (Line 4), resulting in two subgoals corresponding to the
base case and inductive case (Figure~\ref{eg:c}). We then prove the first subgoal with a \textit{bullet} ``\texttt{-}'' (Line 5),
which marks the beginning of the subgoal's proof and causes Coq to
display only this subgoal to the user (Figure~\ref{eg:d}). After
proving this subgoal, we prove the next subgoal with the same bullet symbol (Line 7). These bullets help organize the proof by marking the
beginning of each subgoal and instructing Coq to ensure one subgoal is
proven before moving to the next.

We solve the base case (Figure~\ref{eg:d}) by invoking the
\texttt{auto} tactic (Line 6), which uses symbolic-based proof
automation to discharge simple goals. Next we move on to the second
subgoal---the inductive case. Importantly, this goal includes an
inductive hypothesis in its local context. We first use the
\texttt{simpl} tactic (Line 8), which simplifies the goal by
evaluating the \texttt{+} operator (Figure~\ref{eg:f}). Next, we use
the \texttt{rewrite IHn} tactic (Line 9), which substitutes the
left-hand side of the inductive hypothesis (\texttt{IHn}) in the goal
with its right-hand side (Figure~\ref{eg:g}). Finally, we
\texttt{apply} a previously proven theorem \texttt{plus\_n\_Sm:
  forall\ n\
  m : nat, 1 +
  (n + m) = n + (1
  + m)} from the standard Coq library (Line 10). This theorem
establishes that adding \texttt{1} to the sum of \texttt{n + m} is the
same as adding \texttt{n} to \texttt{m + 1}. A theorem of the form
\texttt{A$\implies$B} can be applied to a goal if its conclusion
(\texttt{B}) matches the current proof obligation, resulting in a new
goal corresponding to its hypothesis (\texttt{A}). The \texttt{apply\
  plus\_n\_Sm} tactic directly solves the current goal, because the
conclusion of \texttt{plus\_n\_Sm} matches and \texttt{plus\_n\_Sm}
has no premises. Since no goals remain, the proof is complete, and we use the \texttt{Qed} command to finish the proof of \texttt{add\_comm}.

\subsection{Hammers}
To facilitate proof construction, Coq is equipped with many established proof automation tactics (e.g., \texttt{auto}). These tactics either completely solve the current goal, or leave it unchanged if they fail. Among them, \textit{hammers}~\cite{coqhammer, sledgehammer, kaliszyk2015hol} are powerful tactics that dispatch goals using external automated
theorem provers (ATPs), such as Vampire~\cite{vampire},
CVC5~\cite{cvc5}, E~\cite{eprover} and Z3~\cite{z3}. Many popular ITPs have hammers, including CoqHammer~\cite{coqhammer} for Coq, SledgeHammer~\cite{sledgehammer} for Isabelle, and
HOLyHammer~\cite{kaliszyk2015hol} for HOL Light.

At a high level, hammers work by first encoding the current goal into
a form solvable by an ATP, typically a formula in first-order logic. This is necessary because ITPs support much richer logic, e.g., higher-order logic, than most ATPs. In order to enable the underlying ATP to use previously proven theorems, a subset of the theorems in the global context are encoded alongside the current goal, the task of selecting a relevant set of these theorems is sometimes called \emph{premise selection}~\cite{irving2016deepmath}. Early hammers relied on heuristics to select premises, while modern hammers typically utilize machine learning algorithms for this purpose. After a goal and the selected premises have been encoded, hammers invoke an ATP. If successful, the proof found by the ATP is translated to a form that can be understood by an ITP. Hammers are typically invoked by applying specific tactics. CoqHammer offers a collection of tactics such as \texttt{hammer}, \texttt{hfcrush}, and \texttt{qsimpl}, each of which automatically proves goals using different strategies. While powerful, hammers only perform a subset of reasoning available to an ITP: they typically do not perform induction, for example. This limits their ability to directly prove complex theorems. Nonetheless, they are effective at accurately solving small subgoals. For instance, the \texttt{hammer} tactic is able to completely solve the goals in Figures~\ref{eg:d} and~\ref{eg:e}, while Coq's \texttt{auto} tactic only solves the first goal.

\section{Formative Study}
\label{errtypes}
While LLMs have previously been used to generate proofs in Coq, they
have not proven particularly effective at the
task~\cite{first2023baldur,
  zhang2023getting, thakur2023copra}. To understand the root causes of
this, we have conducted a formative study to identify the common
errors made by LLMs when asked to generate proof scripts. In this
study, we evaluated the ability of \gpt{}~\cite{gpt35} to prove 579
theorems from Verdi, a distributed system verification
project~\cite{verdi}. Verdi has also been used in other
studies~\cite{yang2019learning, first2020tactok, first2022diversity}.
We carefully designed our prompt based on the widely used retrieval
augmented generation (RAG) method~\cite{lewis2020retrieval}. This
prompt is also used by \name{}, and is described in more detail
in Sections~\ref{retrieval} and \ref{prompt}. We prompted GPT-3.5-turbo-1106 API, the latest version available at the time of this study with the default decoding temperature. For each theorem, we sampled only one proof script. We ran the generated proof in Coq, and recorded the error message of the first encountered error in the proof.

We collected a total of 520 errors and conducted an in-depth manual analysis, following the grounded theory~\cite{glaser2017discovery} and the open coding method~\cite{hancock2001introduction}. The first author first labeled 100 errors and came up with an initial categorization. He then discussed and refined the labels and the categorization with the last author in two meetings. The first author then labeled and categorized
the remaining errors based on the refined labels and categorization.
Finally, all authors met to discuss and finalize the categorization,
where the second author, an expert in theorem proving, offered insights
that further enhanced the categorization. The whole process took
approximately 52 person-hours.
We categorized the 520 errors into seven types, as shown in
Table~\ref{tab:stat_error}.

\begin{table}[htb]
\centering
\resizebox{0.95\columnwidth}{!}{%
\begin{tabular}{|c|c|c|}
\hline
\textbf{Type} & \textbf{\# of occurrences} & \textbf{Percentage} (\%) \\ \hline \hline
\typeFour & 258 & 49.6 \\ \hline
\typeOne & 79 & 15.2 \\ \hline
\typeFive & 61 & 11.7 \\ \hline
\typeTwo & 56 & 10.8 \\ \hline
\typeSix & 44 & 8.5 \\ \hline
\typeThree & 19 & 3.7 \\ \hline
Miscellaneous errors & 3 & 0.6 \\\hline \hline
\textbf{Total }& \textbf{520} & \textbf{100} \\ \hline
\end{tabular}
}
\caption{\label{tab:stat_error} The number of occurrences and percentage of each type of
error.}
\vspace{-20pt}
\end{table}

\textbf{1. \typeFour} (49.6\%): When there is a theorem or hypothesis
of the form \texttt{H: A$\implies$B} and the current goal is
\texttt{B}, the \texttt{apply H} tactic can be used to replace the
goal with \texttt{H}'s premise (\texttt{A}). This tactic requires that
the conclusion of \texttt{H} matches the goal. Attempting to apply a
theorem or hypothesis whose conclusion does not match the goal will
cause the tactic to fail. For example, in
Figure~\ref{coq:apply_error}, the proof state has a hypothesis
\texttt{H: m = n}, which cannot be applied because it does not match
the goal \texttt{n = m}. Applying \texttt{H} fails with the error
message ``\textit{Unable to unify `m=n' with `n=m'}.''
 \begin{figure}[htb]
\centering
\lstset{style=coq}
\begin{lstlisting}
n, m: nat
H: m = n
===========================
n = m
\end{lstlisting}
\vspace{-5pt}
\caption{\label{coq:apply_error}\texttt{apply H} causes a wrong theorem application: ``Unable to
unify `m=n' with `n=m' ''.}
\vspace{-10pt}
\end{figure}

\begin{figure*}[thb]
\includegraphics[width=0.95\textwidth]{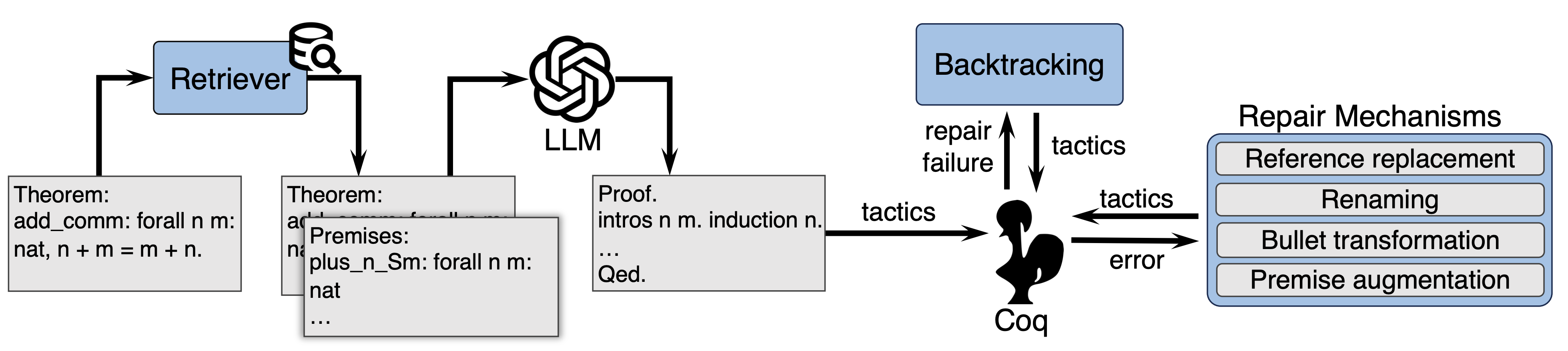}
\vspace{-10pt}
\caption{\label{arch} Overview of \name{}.}
\end{figure*}

\textbf{2. \typeOne} (15.2\%): LLMs can generate incorrect references,
such as a hypothesis that does not exist in the local context or a
theorem that cannot be found in the environment. This is a form of LLM
hallucination~\cite{zhang2023getting}.

\textbf{3. \typeFive} (11.7\%): Given an equation \texttt{Heq}, the \texttt{rewrite Heq} tactic replaces occurrences of the left-hand side
of \texttt{Heq} in the goal with the right-hand side of \texttt{Heq}. The error occurs when a theorem or hypothesis is used for rewriting but its left-hand side does not match any subterms in the goal. For instance, consider the proof state presented in Figure~\ref{coq:rewrite_error}. The \texttt{rewrite H2} tactic fails with the error message ``\textit{Found no subterm matching `b' in the current goal}.''

\begin{figure}[h]
\centering
\lstset{style=coq}
\begin{lstlisting}
a, b, c: nat
H1: a = b
H2: b = c
===========================
a = c
\end{lstlisting}
\vspace{-5pt}
\caption{\label{coq:rewrite_error} \texttt{rewrite H2} fails with: ``Found no subterm matching `b' in the current goal''.}
% \vspace{-5pt}
\end{figure}

\textbf{4.\typeTwo} (10.8\%): The \texttt{intros} tactic is used to
move universally quantified variables and assumptions into the local
context. In some cases, LLMs produce proofs that use \texttt{intros}
to introduce a term with a name that is already in the local context, or
when there is nothing that can be moved to the local context.

\textbf{5. \typeSix} (8.5\%): Some tactics can only be used with specific arguments. This error occurs when a tactic is given an argument that does not satisfy its requirements. For example, \texttt{destruct} and \texttt{induction} can only be applied to arguments with inductive data types such as natural numbers. Both tactics fail with the error ``\textit{Not an inductive product}'' when applied to a non-inductive argument. Conversely, the tactic \texttt{unfold} cannot be applied to arguments with inductive types, and will throw an error ``\textit{Cannot turn inductive into an evaluable reference}.''

Other built-in tactics can be misused in a way specific to the tactic. For example, the \texttt{reflexivity} tactic causes an error when applied to a goal that is not an equality between equivalent terms, while the \texttt{contradiction} tactic fails when the local context does not contain a contradiction.

\textbf{6. \typeThree} (3.7\%): LLMs can generate proofs which misuse
bullets in two ways: (1) the proof tries to proceed to the next goal before the current one is solved, and (2) the proof uses the wrong bullet to focus on a goal. Figure~\ref{coq:bullet_error} illustrates this misuse through two incorrect proofs. In the first proof, the second bullet symbol should be ``\texttt{-}'' instead of ``\texttt{+}'' (Line 6). This leads to an error ``\textit{Wrong bullet \texttt{+}: Expecting \texttt{-}}.'' In the second proof, \texttt{simpl} fails to completely solve the first subgoal (Line 12), so trying to proceed to the second subgoal while the first is unsolved leads to an error ``\textit{Wrong bullet \texttt{-}: Current bullet \texttt{-} is not finished}.''

\begin{figure}[htb]
\centering
\lstset{style=coqline}
\begin{lstlisting}
 Theorem add_comm : forall n m, n + m = m + n.
 (*@\textcolor{gray}{(* wrong proof *)}@*)
 Proof.
   intros. induction n.
   - auto.
   + (*@\textcolor{gray}{(* proof for the inductive case *)}@*)
 Qed.

 (*@\textcolor{gray}{(* wrong proof *)}@*)
 Proof.
   intros. induction n.
   - simpl.
   - (*@\textcolor{gray}{(* proof for the inductive case *)}@*)
 Qed.
\end{lstlisting}
\vspace{-10pt}
\caption{\label{coq:bullet_error} Two examples of bullet misuse.
}
\vspace{-10pt}
\end{figure}

\textbf{7. Miscellaneous errors} (0.6\%) Some errors do not fit into
the previously defined categories. For example, LLMs can generate
special commands like \texttt{Abort}, which terminates a proof without
an error before it is complete. 

A key insight from this formative study is that while LLMs
often generate proof scripts with the right high-level structure, they
often struggle with accurately addressing the sorts of low-level
details that hammers excel at. For example, \gpt{} often knows when to use the \texttt{induction} tactic to decompose theorems into
subgoals, but often fails to generate the right sequence of tactics to prove each subgoal. On the other hand, CoqHammer is good at addressing these subgoals using ATPs. In addition, we found that many proof generation errors are relatively straightforward to fix, e.g., through rule-based transformation, without the need of regenerating the proof from scratch. For instance, both cases of bullet misuse can be repaired by systematically inserting the correct bullet.

\section{Approach}

Guided by our formative study, we propose \name{}, a proof automation approach
that combines LLMs and symbolic methods. Figure~\ref{arch} provides an
overview of \name{}. \name{} includes three components: (1) a
retrieval-augmented proof generation method, (2) a set of repair
mechanisms, and (3) a backtracking procedure.

\subsection{The Overall Algorithm}
Algorithm~\ref{alg:framework} describes the overall generate-then-repair procedure used by \name{}. The inputs are a theorem statement $t$, an environment $env$, and a language model $LM$. First, using the retrieval augmented generation (RAG) method described in Section~\ref{retrieval}, \name{} retrieves relevant premises from $env$ based on $t$ (Line 3). Next, \name{} creates a prompt using $t$ and the selected premises (Line 4), and prompts $LM$ to obtain an initial proof script (Line 5). \name{} then executes these tactics in Coq (Lines 6-15). If an error occurs, \name{} employs a set of repair mechanisms to fix the problem based on the error message, the tactic that throws the error, and the current proof state (Line 9). If \name{} cannot fix an error, it invokes the backtracking procedure (Line 11) described in Algorithm~\ref{alg:back}, which attempts to fix the previous proof using \hammer{}. The proof is successful if no goals remain unsolved after all tactics have been executed (Line 16).

\begin{algorithm}[htb]
  \caption{Framework}
  \label{alg:framework}
  \small{
	\begin{algorithmic}[1]
          \State \textbf{Input}: Theorem statement $t$, Environment
          $env$, Language Model $LM$
    \Function{Prove}{$t$, $env$, $LM$}
    \State $PS\leftarrow$ RetrievePremises($t, env$)
    \State $pt\leftarrow$ BuildPrompt($t, PS$)
    \State $TCS\leftarrow$ $LM$.query($pt$)
    \For{$tc \in \text{TCS}$}
        \State $error \leftarrow$ Coq.execute($tc$)
        \If{$error$}
            \State $repaired\leftarrow$ Repair($error$, $tc$, \textit{current proof state})
            \If{not repaired}
            \State $proof\leftarrow$ Backtracking(\textit{current goal})
                \If{$proof$ is not None}
                    \State Coq.execute($proof$)
                \Else
                    \State \textbf{return} False
                \EndIf
            \EndIf
        \EndIf
    \EndFor
    \State \textbf{return} NoUnsolvedGoals()
    \EndFunction
	\end{algorithmic}
    }
\end{algorithm}

% \vspace{-5 pt}
In the rest of this section, we describe each component of \name{} using the example shown in Figure~\ref{coq:eg_org}. The proof
script shown in the figure was generated by \gpt{} and contains
several errors. The correct proof produced by \name{} is shown in
Figure~\ref{coq:eg_back}.

\begin{figure}[htb]
\lstset{style=coqline}
\begin{lstlisting}
 Lemma sqr_le: forall a: Z, a <= a * a.
 Proof.
   intros. destruct a.
   - reflexivity.
   - induction p.
     + simpl. ring.
     + apply Z_le_dec.
     + apply Z.le_refl.
   - apply Z.eq_le_incl.
 Qed.
\end{lstlisting}
\vspace{-10pt}
\caption{\label{coq:eg_org}A theorem stating $a\leq a\times a$ for any integer $a$ (Line 1), and an erroneous proof (Lines 2 to 10) produced by \gpt{}.}
\end{figure}

\begin{figure}[htb]
\lstset{style=coqline}
\begin{lstlisting}
 Lemma sqr_le : forall a : Z, a <= a * a.
 Proof.
   intros. destruct a.
   - reflexivity.
   - chfcrush use: Zlt_le_succ, Pos2Z.is_pos,
        Z.le_mul_diag_r. 
   - hfcrush.
 Qed.
\end{lstlisting}
\vspace{-10pt}
\caption{\label{coq:eg_back}The correct proof found by \name{}.}
\vspace{-10pt}
\end{figure}

\subsection{Premise Retrieval}
\label{retrieval}
High-quality context is essential for LLMs to produce accurate
responses. For theorem proving, we consider the previously proven theorems and definitions available in the environment as the context of constructing a proof script.

Given there are many available theorems and definitions, it is difficult to encode all of them in the proof generation prompt. Thus, we develop an information retrieval method to identify the ones relevant to the theorem to be proven. Specifically, \name{} predicts relevant premises using Term Frequency-Inverse Document Frequency (TF-IDF)~\cite{sparck1972statistical} and k nearest neighbors (KNN)~\cite{dudani1976distance}.
While more advanced methods such as deep learning~\cite{irving2016deepmath} can be more accurate, they also require significant amounts of training data and can take a longer
time to make predictions~\cite{farber2015random, kuhlwein2012overview}. The premises predicted by the KNN algorithm are initially ranked by their TF-IDF scores. \name{} then employs the BM25 algorithm~\cite{bm25wiki} to rerank these premises based on their text similarity to the statement of the theorem, since we observed that BM25 tends to rank premises used in human-written proofs higher than TF-IDF.

\vspace{-5pt}

\subsection{Prompt Design}
\label{prompt}
\begin{figure}[htb]
\includegraphics[width=0.45\textwidth]{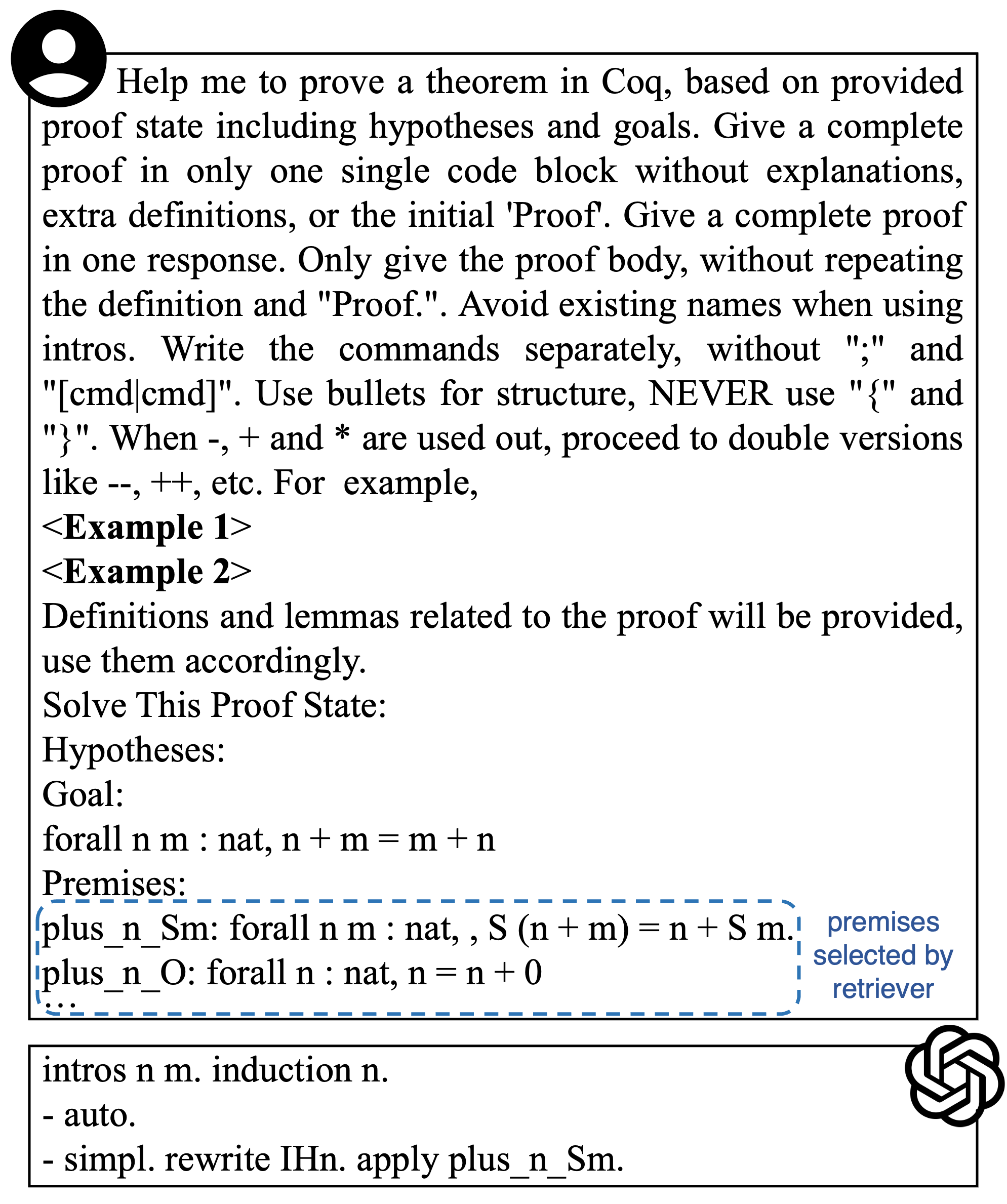}
\vspace{-10pt}
\caption{\label{coq:prompt} An example of the prompt constructed by \name{}, and the response of \gpt{}.}
\vspace{-10pt}
\end{figure}

To optimize the quality of the initial proof generated by the LLMs, we
carefully designed the prompt used by \name{} following strategies for
few-shot in-context learning. Our strategy for designing this prompt
was inspired by recent findings that LLMs can produce instructions
that are superior or equivalent to those crafted by humans~\cite{APE}.
We first asked GPT-4 to infer the five most effective instructions for
two theorems accompanied by human-written proof scripts. Next, we
constructed candidate prompts by combining these five sets of
instructions with the two examples, premises and a new theorem to be
proven. The inclusion of the two example theorems in this query was
meant to demonstrate the correct Coq syntax and our desired proof style.
The first author then manually examined 20 proofs produced by the LLM
in response to these prompts and chose the prompt that yielded the
highest quality proofs. Proof quality was assessed based on
correctness and adherence to the instructions, e.g., using bullets for structure, etc. When multiple proofs met these criteria, the simplest (i.e.
shortest) correct proof was preferred. Figure~\ref{coq:prompt} illustrates the final prompt template with an example and the response of GPT-3.5.

\vspace{-10pt}

\subsection{Repair Mechanisms}
\label{repair}
The proofs produced by LLMs typically feature a good high-level structure that decomposes the proof into reasonable subgoals. However, most of these proofs are rejected by Coq due to errors, as discussed in Section~\ref{errtypes}. To address this issue, we have developed a set of repair mechanisms to handle the common types of errors.

\textbf{\repairOne} When LLMs generate tactics referencing undefined
theorems or hypotheses, \name{} systematically searches in the local
and global context for theorems and hypotheses with similar names, in
order to find suitable replacements. Specifically, \name{} first
collects a set of candidates, including the relevant theorems selected by
the retrieval method, and hypotheses in the local context. It then
ranks these candidates using BM25 based on the similarity of their
names to the initial undefined reference name. Then, \name{} iteratively replaces the undefined reference name in the tactic with each candidate and asks Coq to re-execute the updated tactic, until the the tactic succeeds. For example, if a candidate proof uses the tactic \texttt{apply in\_remove\_all} but \texttt{in\_remove\_all} does not exist, \name{} searches for similar reference names. It first ranks the selected theorems and hypotheses based on the similarity of their names to \texttt{in\_remove\_all}. Then, \name{} iteratively replaces \texttt{in\_remove\_all} with the candidates and eventually finds a tactic \texttt{apply in\_remove\_all\_preserve}, which solves the goal.

\textbf{\repairTwo} If a proof script tries to introduce a term using \texttt{intros} but the specified name already exists in the local context, \name{} appends an apostrophe to the specified name and updates the tactic accordingly. For example, if the tactic \texttt{intros H} is used but \texttt{H} already exists in the local context as a hypothesis, \name{} updates the tactic to \texttt{intros H'}.
If there is nothing that can be moved to the local context, \name{}
simply drops the \texttt{intros} tactic from the current proof script.

\textbf{\repairThree} \name{} handles bullet misuse in two ways,
depending on the two specific scenarios described in Section 3. First, if the current goal has been
solved and the next goal is focused on using the wrong bullet, Coq will
indicate the expected bullet, and \name{} will simply update the proof
to use it. Second, if the proof attempts to proceed to the next goal or
finish the proof while there are still unsolved goals, \name{} will delegate the repair effort to the backtracking procedure described in the next section.

To illustrate this, consider the last subgoal produced by
\texttt{destruct a}, as shown in Line 9 of Figure~\ref{coq:eg_org}.
The \texttt{apply Z.eq\_le\_incl} tactic fails to fully solve this
subgoal, so attempting to finish the proof with \texttt{Qed} (Line 10)
causes an error. To fix this, \name{} starts the backtracking
procedure using the goal that results from the \texttt{apply
  Z.eq\_le\_incl} tactic. Eventually, the backtracking procedure
replaces the \texttt{apply Z.eq\_le\_incl} tactic with the
\texttt{hfcrush} tactic (Line 6 in Figure~\ref{coq:eg_back}) and completely solves the goal.

\textbf{\repairFour} LLMs can produce proof scripts that misuse a theorem, resulting in a \textit{wrong theorem application}, \textit{wrong rewriting}, or \textit{tactic misuse} error. Despite this, the misused theorems are still potentially helpful: although used improperly in the proof script, they might still aid in solving the goal if used in a different manner. Based on this insight, \name{} leverages \hammer{} to determine how to use these theorems correctly. Specifically, it employs the \texttt{qsimpl} tactic provided by \hammer{}, which accepts a list of theorems as arguments. \texttt{qsimpl} uses sophisticated heuristics to identify which theorems can be applied and simplifies the current goal accordingly. Similar tactics are available in other proof assistants. \name{} executes \texttt{qsimpl} with a misused theorem as the argument, allowing it to automatically discover the correct usage of this theorem. For example, if a tactic \texttt{apply Zlt\_le\_succ} causes an error because its conclusion does not match the current goal, \name{} will execute \texttt{qsimpl use: Zlt\_le\_succ} to utilize this theorem despite its initial misuse. 

\subsection{Backtracking}
\label{backtrack}
\name{} leverages \hammer{} to solve goals that the initial script fails to prove due to errors that cannot be repaired.  Although other proof automation techniques could be employed, we found hammers to be effective in practice. Applying a wrong tactic can result in a new goal that is more difficult, or even impossible to prove. Thus, when \hammer{} fails to solve a goal, it is clear that \name{} needs to backtrack to an earlier point in the proof to see if it can be solved instead. Particularly, if a tactic produces multiple subgoals, all these subgoals must be proven. If \name{} fails to prove any of them, it needs to revert to the goal before that tactic. For example, when reasoning by induction, if the base case is proven but the inductive case fails, the entire induction attempt has failed, and we need to try a different proof strategy instead of induction.

Algorithm~\ref{alg:back} presents our backtracking procedure, which aims to prove unsolved goals using \hammer{}. The input to the procedure is an unsolved goal $g$. If $g$ is successfully solved by \hammer{}, it returns the proof found by \hammer{} immediately (Lines 4-5). Otherwise, \name{} reverts to the goal before the last applied tactic, and tries \hammer{} again. If the last command is a bullet (Line 6), it means that our algorithm will not be able to prove this subgoal using \hammer{}. When this happens, the algorithm identifies the tactic that produced this subgoal as the $root$ (Line 7), and then discards $root$ and all its associated subgoals (Line 8). Having the LLM structure its proof using bullets helps PALM identify which parts of the proof need to be dropped when a subgoal fails. If the last tactic is not a bullet, \name{} simply reverts to the goal before that tactic (Line 10). This loop continues until \hammer{} succeeds (Line 5) or no tactics remain in which case the repair attempt fails (Line 3).
\begin{algorithm}[!t]
  \caption{Backtracking}
  \label{alg:back}
  \small{
	\begin{algorithmic}[1]
    \State \textbf{Input}: Unsolved Goal $g$
    \Function{Backtrack}{$g$}
        \While{exist tactics}
            \If{$g$ is solved by CoqHammer}
                \State \textbf{return} CoqHammer.getProof()
            \ElsIf {the last tactic is a bullet}
                \State $root\leftarrow$ the tactic produces this subgoal
                \State discard $root$ and its subgoals
            \Else
                \State Coq.undo()
            \EndIf
        \EndWhile
    \State \textbf{return} None
    \EndFunction
	\end{algorithmic}
    }
\end{algorithm}

We demonstrate the backtracking procedure on the \texttt{induction p} tactic and its subgoals (shown in Lines 5-8 of
Figure~\ref{coq:eg_org}). Initially, all tactics up to \texttt{ring}
(Line 6) are executed without errors. However, the \texttt{ring}
tactic fails and cannot be repaired, so \name{} starts backtracking.
The input is the goal resulting from the execution of the last tactic,
\texttt{simpl}. Algorithm~\ref{alg:back} invokes \hammer{}, but it
fails to solve the goal, so \name{} reverts the \texttt{simpl} tactic
and invokes \hammer{} again, but \hammer{} still fails. At this point,
the algorithm hits a bullet (``\texttt{+}''), and there are no
remaining tactics that can be repaired using \hammer{}. This indicates that
the first subgoal produced by \texttt{induction p} cannot be proven,
leading to the failure of the entire induction attempt. Accordingly,
\name{} discards the \texttt{induction p} tactic (Line 5) along with
the tactics corresponding to all its subgoals (Lines 6-8).
Algorithm~\ref{alg:back} then reverts to the second subgoal produced
by \texttt{destruct a}, which is successfully solved by \hammer{}. The proof found by \hammer{} is presented in Lines 5-6 of Figure~\ref{coq:eg_back}.

\begin{figure}[htb]
\includegraphics[width=0.45\textwidth]{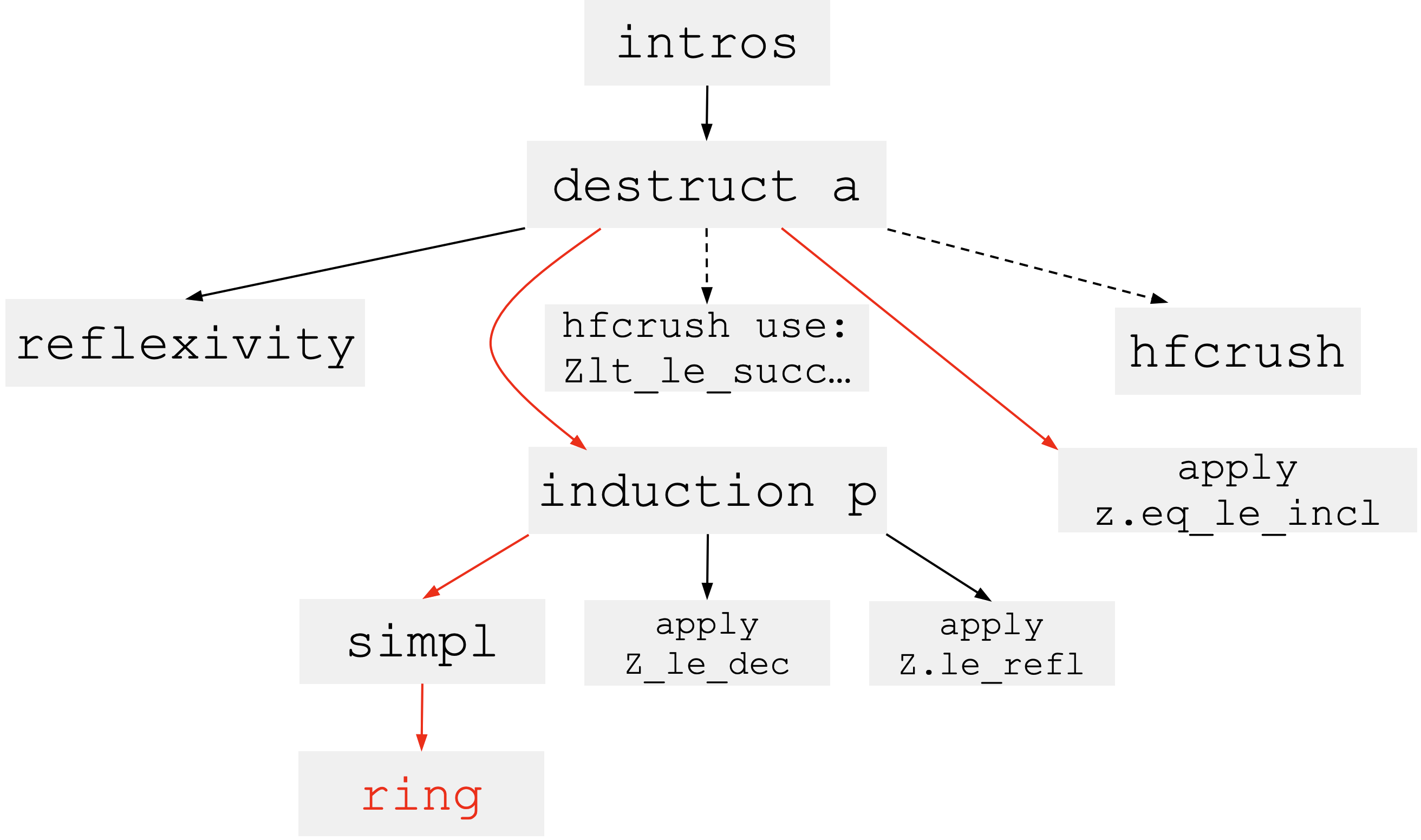}
\caption{Visualization of our backtracking repair algorithm. The
  red lines indicate the reverted tactics, and the dashed lines
  indicate the tactics found by \hammer{} during backtracking.}
\label{coq:eg_tree}
\end{figure}

\section{Evaluation}
Our experimental evaluation of our approach addresses four key
research questions:

\begin{itemize}
\item \textbf{RQ1:} Is \name{} more effective at proving theorems
  than other state-of-the-art proof automation approaches? %
\item \textbf{RQ2:} Can \name{} generalize to other LLMs with
  different parameter sizes? %
\item \textbf{RQ3:} How much does each component of \name{} contribute
  to its effectiveness? %
 \item \textbf{RQ4:} Is \name{} time-efficient? 
 % Can \name{} generate proofs in a reasonable amount of time?
\end{itemize}

We conducted experiments on a workstation with an AMD EPYC 7313
CPU, an NVIDIA A5500 GPU, and 512GB memory. The operating system was
64-bit Ubuntu 22.04 LTS.

\subsection{Comparison baselines}
We compare \name{} against three state-of-the-art proof automation
approaches: \passport{}~\cite{sanchez2023passport},
\proverbot{}~\cite{sanchez2020generating} and \dspfull{}~\cite{jiang2022draft}. Both \passport{} and \proverbot{}
are machine learning methods. \passport{} employs a
Tree-LSTM~\cite{tai2015improved} to model proof states, incomplete
proof scripts, and identifiers in proofs. \proverbot{} adopts an RNN
to model manually engineered features of the proof states. \dsp{}
prompts LLMs to translate natural language proofs into formal proof sketches that outline high-level proof steps without low-level
details. The informal proofs can be either written by humans or generated
by LLMs. It then uses off-the-shelf proof automation tools such as
hammers to fill in the gaps. Unlike \dsp{}, \name{} does not require
informal proofs, and employs repair mechanisms and a backtracking
procedure to fix proof errors. As human-written proofs were unavailable
for the benchmarks used in our test set, in order to reproduce \dsp{},
we used \gpt{} to generate informal proofs and sketches, and use
CoqHammer as the underlying proof automation tool.

\subsection{Benchmark construction}
Following prior work~\cite{yang2019learning, sanchez2023passport, sanchez2020generating}, we use the test set of CoqGym~\cite{yang2019learning} as the evaluation dataset, which consists of 13,137 theorems from 27 open-source Coq projects. Since the theorems from the Verdi project used in our formative study are also included in CoqGym, we exclude them to avoid biases.
% Since our repair mechanisms are based on errors observed in this study, we drop these benchmarks from our evaluation to prevent overfitting. % CoqGym builds the projects using Coq 8.9, and extracts % theorems and proofs from them.
As we ran the baselines on CoqGym, we found that \passport{} is compatible exclusively with Coq 8.9, and relies on CoqGym's original dataset. \proverbot{}, which does not use CoqGym, supports only newer versions of Coq, namely Coq 8.10, 8.11, and 8.12. To ensure fairness, our evaluation is conducted on a subset of CoqGym, including \totalnum{} theorems that are compatible with all relevant versions of Coq. We implement \name{} for Coq 8.10, 8.11, and 8.12, since many language features and standard libraries of Coq 8.9 are outdated~\cite{coqchangelog}.

\subsection{Results}
In RQ1, we compare the performance of \name{} using \gpt{} as the underlying LLM against the baselines. In RQ2, we evaluate the performance of \name{} when using different LLMs.

\subsubsection{RQ1: Effectiveness of \name{}}
Table~\ref{eval:table_baseline} shows the number and percentage of theorems each approach can successfully prove. Compared with existing approaches, \passport{}, \proverbot{}, and \dsp{}, \name{} proves \outpassport{}, \outproverbot{}, and \outdsp{} more theorems, respectively.
Since \passport{} and \proverbot{} use less powerful LSTM and RNN models, they prove fewer theorems than \name{} and \dsp{} which leverage LLMs.
\dsp{} underperforms \name{} due to two reasons. First, we use \gpt{} to generate informal proofs needed by \dsp{}, but it may introduce errors in the generation process. Second, since \dsp{} does not repair errors, any errors in the generated proof, no matter how big or small the errors are, will lead to a proof failure. Compared with \dsp{}, \name{} repairs errors in the proofs generated by the LLM, and performs backtracking to regenerate previous proof steps when hammers fail.

\begin{table}[]
\begin{center}
\begin{tabular}{lcc}
\hline
\textbf{Approach} & \textbf{\# of Theorems Proven} &  \\ \hline
\passport{} & \succnumpassport{} (\succratepassport{}) & \\
\proverbot{} & \succnumproverbot{} (\succrateproverbot{}) & \\
\dspfull{} & \succnumdsp{} (\succratedsp{}) & \\
\hline
\gpt{} & \succnuminitial (\succrateinitial)\\
\scalebox{1.2}{$+$} \name{} & \succnum{} (\succrate{}) \\
\hline
\gptfour{} & \succnuminitialfour (\succrateinitialfour) \\
\scalebox{1.2}{$+$} \name{} & \succnumfour{} (\succratefour{}) \\
\hline
\llamaseven{} & \succnuminitialllamaseven (\succrateinitialllamaseven) \\
\scalebox{1.2}{$+$} \name{} & \succnumllama{} (\succratellama{}) \\
\hline
\llamaeight{} & \succnuminitialllamaeight (0.1\%) \\
\scalebox{1.2}{$+$} \name{} & \succnummistral{} (\succratemistral{}) \\
\hline
\end{tabular}
\caption{Theorems proved by each approach.}
\label{eval:table_baseline}
\end{center}
\vspace{-10pt}
\end{table}

\begin{figure}[htb]
\includegraphics[width=0.45\textwidth]{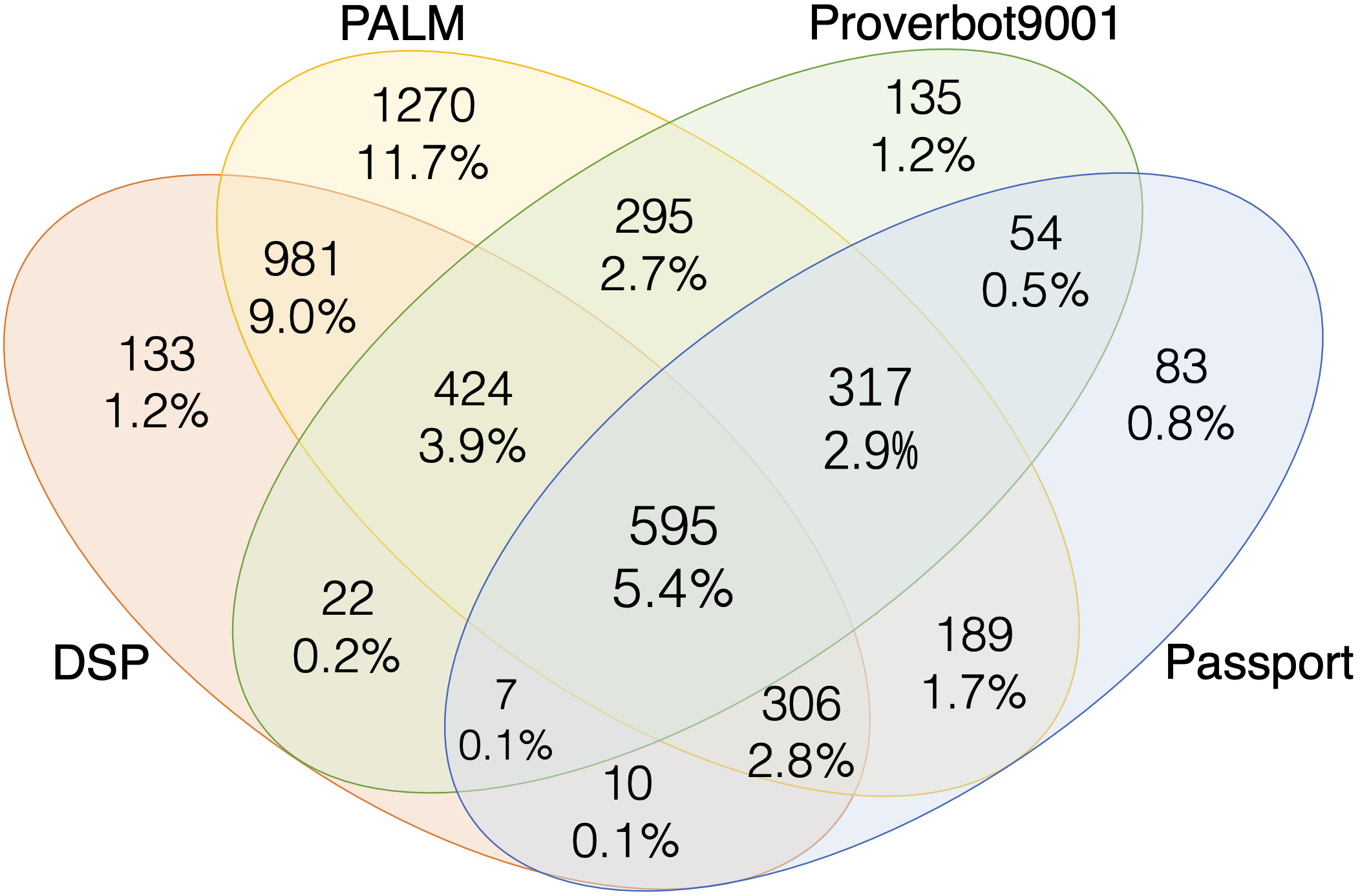}
\vspace{-10pt}
\caption{\label{eval:venn} Breakdown of theorems proven by each combination of approaches.}
%\vspace{-10pt}
\end{figure}

Figure~\ref{eval:venn} presents a Venn diagram illustrating the theorems proven by each approach. All four approaches can collectively prove 4821 (44.5\%) theorems, of which only 444 cannot be proven by \name{}. The three baselines are able to prove \succnumcombined{} distinct theorems in total, and \name{} outperforms their combination by \outcombined{}. Moreover, \name{} proves 1270 theorems that none of the other approaches can prove.

\begin{figure}[htb]
\includegraphics[width=0.45\textwidth]{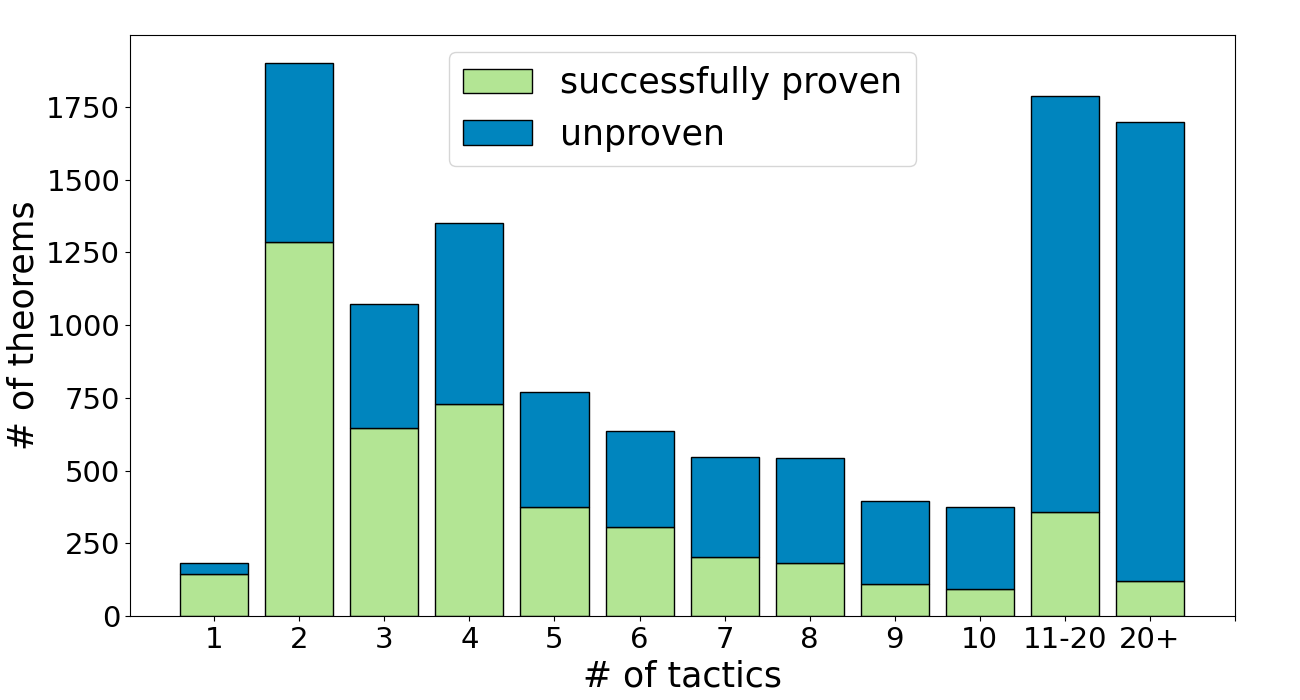}
\vspace{-10pt}
\caption{\label{eval:length} Distribution of theorems that are proven or not proven by \name{}, categorized by the number of tactics in the ground-truth proofs.}
\end{figure}

We further analyze the complexity of the theorems that \name{} proves,
using the number of tactics in a proof as a proxy metric for theorem
complexity. Figure~\ref{eval:length} shows the distribution of
theorems that are proven or not proven by \name{}, categorized by the
number of tactics in the ground-truth proofs. The average number of
tactics in the ground-truth proofs is 5.84 and the median is 4,
suggesting \name{} is more effective with simpler proofs. 
Moreover, \name{} can prove 129 theorems that require 20 tactics or
more, outperforming \passport{} (11), \proverbot{} (30) and \dsp{}
(48).

% \vspace{-20pt}

\finding{Finding 1: \name{} is more effective than \passport{}, \proverbot{} and \dsp{} on our benchmarks, proving significantly more
  theorems. Notably, \name{} proves 1270 theorems that none of the
  other approaches can prove. Additionally, \name{} can prove a larger
  number of complex theorems than other approaches.}

\subsubsection{RQ2: Generalizability of \name{}}
To demonstrate the generalizability of \name{} across LLMs with different parameter sizes, we further evaluate \name{} with GPT-4o~\cite{gpt4o}, Llama-3-70B-Instruct~\cite{llama3} and Llama-8B-Instruct~\cite{llama3} as the underlying LLMs.

Table~\ref{eval:table_baseline} presents the theorems proven by each LLM individually, and by \name{} when using them as underlying LLMs. We observe that all evaluated LLMs perform poorly when used alone, proving only 0.1\%-6.4\% of theorems. Augmenting these LLMs with \name{} significantly improves the performance. With the most powerful GPT-4o model, \name{} proves \succnumfour{} theorems, achieving a 5.5\% absolute improvement compared with using the second most powerful LLM, \gpt{}. This highlights the potential enhancements \name{} can achieve with the latest LLMs. When using Llama-3-70B-Instruct, \name{} proves \succnumllama{} theorems, which is comparable with the result obtained using \gpt{}. When using the smaller Llama-8B-Instruct, \name{} proves \succnummistral{} theorems, 21.6\% fewer than when using \gpt{}. Despite this, \name{} still outperforms \dsp{} by 38.5\%, suggesting it can be effective even when using less powerful LLMs. Using all four LLMs, \name{} successfully proves a total of 5210 theorems.

\finding{Finding 2: \name{} generalizes to other LLMs of different
  parameter sizes, and performs better when using larger LLMs.}

\subsubsection{RQ3: Effectiveness of each component} We have conducted an
ablation study to evaluate the effectiveness of each component within \name{}.

\textbf{Effectiveness of the repair mechanisms.} To study the
effectiveness of each repair mechanism, we constructed four variants
of \name{}: \varnoref{}, \varnoname{}, \varnobullet{}, and
\varnoaug{}. These variants disable the reference replacement,
renaming, bullet transformation, and premise augmentation mechanisms,
respectively. Table~\ref{eval:table_ablation} presents the evaluation
results of \name{} and the variants.

Overall, \name{} consistently proves \outnoref{}-\outnoaug{} more theorems than each variant, indicating the importance of each of our repair mechanisms. Furthermore, all variants continue to prove 65.2\%-71.5\% more theorems than \dsp{}, demonstrating that \name{} remains effective even when equipped with partial repair mechanisms.

\finding{Finding 3: Each repair mechanism of \name{} contributes to
  its ability to prove theorems.}

\textbf{Effectiveness of backtracking.} We evaluated the effectiveness
of the backtracking procedure (Algorithm~\ref{alg:back}) by constructing an additional variant, called \varnobacktrack{}. It does not perform backtracking when it fails to repair an error, and immediately terminates the proof process instead. As shown in Table~\ref{eval:table_ablation}, \name{} significantly outperforms \varnobacktrack{} by \outnobacktrack{}, indicating that the backtracking procedure is essential to proving many theorems.

\finding{Finding 4: The backtracking procedure is essential to
  \name{}'s effectiveness, enabling it to prove $5\times$ more
  theorems than only utilizing the repair mechanisms.}

\textbf{Effectiveness of our premise retriever.} To investigate the effectiveness of our premise retriever, we constructed a variant called \varnoretriever{}, which does not add any premises to the proof generation prompt.

Table~\ref{eval:table_ablation} shows that \name{} outperforms
\varnoretriever{} by \outnoretriever{}, which underscores that the
premise retriever enables LLMs to produce higher-quality proof
scripts.

\begin{table}[]
\begin{center}
\begin{tabular}{lcc}
\hline
\textbf{Technique Variant} & \textbf{\# of Theorems Proven} &  \\ \hline
\varnoref{} & \succnumnoref{} (\succratenoref{}) & \\
\varnoname{} & \succnumnoname{} (\succratenoname{}) & \\
\varnobullet{} & \succnumnobullet{} (\succratenobullet{}) & \\
\varnoaug{} & \succnumnoaug{} (\succratenoaug{}) & \\
\hline
\varnobacktrack{} & \succnumnobacktrack{} (\succratenobacktrack{}) & \\
\hline
\varnoretriever{}{} & \succnumnoretriever{} (38.2\%) & \\
\hline
\namegpt{} & \succnum{} (\succrate{}) \\
\hline
\end{tabular}
\caption{Effectiveness of each \name{} component.}
\label{eval:table_ablation}
\end{center}
\vspace{-20pt}
\end{table}

\finding{Finding 5: The premise retriever is useful to \name{}, helping it to prove \outnoretriever{} more theorems.}

\subsubsection{RQ4: Efficiency of \name{}}
On average, \name{} takes 32.89 seconds to successfully prove a
theorem, while \passport{}, \proverbot{} and \dsp{} require 3.1, 4.7
and 8.2 seconds, respectively. The main source of time overhead for
\name{} is its use of \hammer{}. On average, \hammer{} is invoked 1.96
times per proof, with each invocation having a timeout of 10 seconds.
This additional time is justified by \name{}'s ability to prove more
complex theorems than other approaches. We further examined the
time each approach takes on all theorems, regardless of whether they
were successfully proven or not. On average, \name{} takes 105.6
seconds, while \passport{}, \proverbot{} and \dsp{} take 67.2, 31.8
and 20.6 seconds respectively. Additionally, each successful CoqHammer
invocation averages 4.3 seconds, with 79.3\% of successful invocations
completing in under 5 seconds. This suggests that \name{} could prove
a substantial number of theorems even with a shorter CoqHammer time
limit, while a longer limit would potentially benefit more complex
proofs.

\finding{Finding 6: On average, \name{} takes longer to prove theorems
  than other approaches, but this overhead is acceptable
  given that \name{} proves more complex theorems.}

\subsection{Case Studies}
Despite its effectiveness, \name{} still fails to prove 59.6\% theorems in
our dataset. To understand the underlying reasons for these failures,
we randomly sampled 100 theorems that \name{} fails to prove and
conducted a manual analysis. Table~\ref{eval:table_case} outlines the
3 primary reasons for these failures.\footnote{The columns in Table~\ref{eval:table_case} sum to more than 100\% because a single theorem can fail for multiple reasons.} To illustrate these reasons further, we now describe a typical case of failure for each.

\begin{table}[htb]
\begin{center}
\begin{tabular}{lcc}
\hline
\textbf{Reason} & \textbf{\# occurrences} &  \\ \hline
Premises not retrieved & 58 (58\%) & \\
Premises retrieved but not used & 14 (14\%) & \\
Tactics not used & 39 (39\%) & \\
\hline
\end{tabular}
\caption{The reasons causing \name{} to fail.}
\label{eval:table_case}
\end{center}
\vspace{-25pt}
\end{table}

\subsubsection{Missing premises}
% TBH,
A key reason for \name{}'s failures (58\%) is the omission of necessary
premises in the retrieval process.
Figure~\ref{coq:case_premise_not_retrieved} presents a theorem that
\name{} fails to prove because a critical premise,
\texttt{reduceplus\_cb1}, was not retrieved. Hence this premise cannot
be used by the LLM, hindering the proof process.

\begin{figure}[htb]
\lstset{style=coq}
\begin{lstlisting}
Theorem reducestar_cb1 :
 forall (a : poly A0 eqA ltM) (b : list (Term A n))
   (Q : list (poly A0 eqA ltM)),
 reducestar A A0 A1 eqA invA minusA multA divA eqA_dec n ltM ltM_dec Q
   (s2p A A0 eqA n ltM a) b -> CombLinear (a :: Q) b.

  (*@\textcolor{gray}{(* Human written proof *)}@*)
  intros a b Q H'; inversion H'; auto.
  apply reduceplus_cb1; auto.

  (*@\textcolor{gray}{(* LLM generated proof *)}@*)
  intros a b Q Hred. induction Hred.
  - constructor. - apply CombLinear_1; auto.
\end{lstlisting}
\vspace{-5pt}
\caption{\label{coq:case_premise_not_retrieved}
A failure case~\cite{casebuchberger} because \texttt{reduceplus\_cb1} is not retrieved.}
% \vspace{-15pt}
\end{figure}
\subsubsection{Premises retrieved but not used}
In 14\% of the failures, even when a premise is successfully retrieved and
included in the prompt, it may not be used by the LLM.
Figure~\ref{coq:case_premise_not_used} shows a case where the lemmas
\texttt{map\_insert} and \texttt{map\_map\_exchange} are included in
the prompt, but they are not used by the LLM, causing \name's failure
to prove the theorem. Although providing CoqHammer with unused
retrieved premises during the backtracking process could solve such
issues, we choose not to do so, as providing too many unrelated
premises slows down \hammer{} and can lead it to timeout.
% from the \textit{dblib} project

\begin{figure}[htb]
\lstset{style=coq}
\begin{lstlisting}
Lemma map_insert_map:
  forall A (f g h : A -> A) x (a : A) e,
  (forall a, f (g a) = g (h a)) ->
  map f (insert x a (map g e)) =
  insert x (f a) (map g (map h e)).

  (*@\textcolor{gray}{(* Human written proof *)}@*)
  intros. rewrite map_insert. f_equal.
  eapply map_map_exchange. eauto.

  (*@\textcolor{gray}{(* LLM generated proof *)}@*)
  intros. apply map_insert_eq. apply H.
\end{lstlisting}
\vspace{-5pt}
\caption{\label{coq:case_premise_not_used}
A failure case~\cite{casedblib} where the LLM does not use \texttt{map\_insert} and \texttt{map\_map\_exchange} provided in the prompt.}
\vspace{-2pt}
\end{figure}

\subsubsection{Tactics not used}
Some theorems require specific tactics to be proven, and \name{} will
fail if these tactics are not included in the proof script generated
by the LLM. This accounts for 39\% of the failure cases. Figure~\ref{coq:case_tactic_not_used} shows an example where the proof of a theorem requires the use of the \texttt{induction} tactic. Since the initial proof script did not include this tactic, and both \hammer{} and our repair mechanisms do not perform induction, \name{} cannot prove this theorem.

\begin{figure}[htb]
\lstset{style=coq}
\begin{lstlisting}
Lemma last_holder'_unlock_none : forall tr h c,
  last_holder' h tr = Some c ->
  slast_holder' h (tr ++ [(Client c, inl Unlock)])=None.

  (*@\textcolor{gray}{(* Human written proof *)}@*)
  induction tr; intros; simpl in *; repeat break_match; intuition. congruence.

  (*@\textcolor{gray}{(* LLM generated proof *)}@*)
  intros tr h c i n H1 H2 H3 H4.
  apply (last_holder'_no_out_inv tr h (Client c) n).
  apply H1.

\end{lstlisting}
\vspace{-5pt}
\caption{\label{coq:case_tactic_not_used}
A failure case~\cite{caseverdi} where the LLM does not perform induction.}
\vspace{-10pt}
\end{figure}

\section{Discussion}
\subsection{Threats to Validity}
\noindent \textbf{Internal validity.} One threat to internal validity
comes from the inherent randomness of LLMs. This randomness is due to
the use of temperature sampling~\cite{zhu2024hot, ackley1985learning}
as the decoding strategy, where LLMs randomly select the next token
based on a probability distribution. To reduce this threat, we conduct
large-scale experiments using three state-of-the-art and widely used LLMs:
\gpt{}, \gptfour{}, Llama-3-70B-Instruct, and Llama-3-8B-Instruct, as the underlying LLMs for \name{}. We evaluate their performance across a benchmark consisting of \totalnum{} theorems from diverse domains. The consistent results observed in our experiments help reduce this threat. Another threat is that due to the limitation of computational resources and evaluation time, we only run each of our experiments once; this may introduce statistical biases into our results.

\noindent \textbf{External validity.} The threat to external validity
is alone the generalizability of our experimental results. We implement and evaluate only on Coq, while other widely used ITPs, such as Isabelle, HOL Light, and Mizar, are not included. Nonetheless, we believe the approach and algorithm in \name{} can be easily applied to other ITPs that use tactics for proof construction and support automation tools like CoqHammer. However, the specifics will need to be adapted for different tactic languages. For example, Isabelle structures subgoals using the `case` keyword instead of bullets, thus the bullet transformation needs to be modified. We plan to extend \name{}'s implementation to other ITPs, and evaluate it across more diverse datasets in our future work.

\noindent \textbf{Construct validity.} One potential threat to
construct validity is that we use the number of tactics in ground-truth proofs as a proxy metric for the theorem complexity. This metric
may not accurately reflect the real complexity of a theorem.

\subsection{Limitations and Future Work}
\name{} fundamentally depends on the initial proof script generated by LLMs. If the LLM generates a completely wrong initial proof, \name{}
struggles to fix it. Future improvements to \name{} could involve
leveraging LLMs to repair incorrect proofs~\cite{first2023baldur} or
sampling multiple initial proofs.

The initial proof script sometimes fails to use relevant tactics,
such as a user-defined tactic with an ambiguous or uninformative name.
As a result, \name{} cannot effectively prove theorems that depend on
custom user-defined tactics. This can be improved by adopting more
powerful retrievers~\cite{yang2024leandojo} that learn from the usage
patterns of these user-defined tactics.

Finally, we did not spend significant effort optimizing the prompt used by \name{}, since our focus was not on prompt engineering. Different combinations of instructions or using a more advanced prompting design, such as Chain-of-Thought~\cite{wei2022chain} and Least-to Most~\cite{zhou2022least} prompting, may improve the performance of \name{}. These approaches are worth exploring in future work.
\section{Related Work}

\paragraph{Machine Learning for Formal Verification}

There have been various machine learning-based techniques that aim to
automatically generate formal proofs for different ITPs.
ASTactic~\cite{yang2019learning} is the first deep learning-based
proof generation technique for ITPs. It leverages
Tree-LSTM~\cite{tai2015improved} to model proof states with all
Coq terms parsed into abstract syntax trees (ASTs), and searches for a
complete proof via depth-first search (DFS). Many other techniques
have been proposed to enhance the performance of ASTactic.
TacTok~\cite{first2020tactok}, for example, models not only the proof
states, but also the incomplete proof scripts to provide more context
information. To enlarge the search space,
Diva~\cite{first2022diversity} combines multiple models that are
trained with different hyperparameters, such as learning rate and
embedding size, and different orderings of training data.
Passport~\cite{sanchez2023passport} further extends ASTacic and TacTok by adding new encoding mechanisms for
identifiers in proof scripts. These techniques are all evaluated on
the CoqGym~\cite{yang2019learning} dataset.
Proverbot9001~\cite{sanchez2020generating} learns to predict the
tactics and arguments using an RNN model and a set of manually
engineered features. It also leverages advanced search algorithms such as A-star, and several pruning techniques.

Unlike existing machine learning methods that require significant training, \name{} leverages LLMs and does not require any training or fine-tuning. Instead of using search strategies, \name{} employs repair mechanisms and a backtracking procedure to address errors and solve the goals that LLMs fail to prove.

\vspace{-5pt} \paragraph{Language Models for Formal Verification}
Recently, there has been considerable interest in applying LLMs to
formal verification. The most related work is Draft, Sketch, and Prove
(DSP)~\cite{jiang2022draft}. Similar to \name{}, DSP also synergizes LLMs and automated theorem provers. DSP uses LLMs to
translate natural language proofs (i.e., informal proofs) into formal proof sketches that outline high-level steps without low-level details. Then, it uses off-the-shelf proof automation tools such as hammers to fill in the gaps. In contrast, \name{} does not require informal proofs to guide the generation of machine-checked proofs. While DSP reports a failure once proof automation tools cannot fill in a gap, \name{} employs a backtracking procedure to regenerate previous proof steps when hammer fails. Additionally, \name{} adopts repair mechanisms to address common errors made by LLMs.

Minerva~\cite{lewkowycz2022solving} is a LLM trained on mathematical
datasets and achieves state-of-the-art performance on quantitative
reasoning tasks. Baldur~\cite{first2023baldur} fine-tunes Minerva to
create (1) a \textit{proof generation model} that generates whole
proofs given a theorem, and (2) a \textit{proof repair model} that
repairs an incorrect proof given the error message. To train the
proof generation model, it constructs a dataset by
concatenating the proof steps of each theorem from the PISA
dataset~\cite{jiang2021lisa}. The PISA dataset consists of 183K
theorems collected from the Isabelle standard
library~\cite{nipkow2002isabelle} and the Archive of Formal
Proofs~\cite{AFP}. To train the proof repair model, it
samples from the proof generation model for each theorem in
PISA, and records the error messages returned from the ITP for each
erroneous proof. The dataset comprises tuples of incorrect proofs,
error messages, and correct proofs. Compared with Baldur, \name{}
adopts error-specific repair mechanisms to effectively address the
errors. Although Baldur does not perform any search, it needs 64
samples for proof generation and 32 samples for proof repair to achieve high performance, while \name{} only samples once from LLMs.
We have not compared the performance of \name{} with Baldur because
the Minerva model used by Baldur is not open-sourced, and reproducing Baldur by fine-tuning publicly accessible LLMs such as Llama-3 would require extensive computational resources. For instance, the proof generation model of Baldur is fine-tuned on 64 Google TPU v3, with a total of 1024 GB memory. To fine-tune Llama-3-8b with the same settings, over 512 GB GPU memory (around 7 A100s) is required. Furthermore, the proof repair dataset of Baldur consists of 150K tuples of wrong proofs, error messages, and correct proofs in Isabelle. To extend Baldur for Coq, a similar dataset would need to be constructed for Coq.

Thor~\cite{jiang2022thor} augments the PISA
dataset~\cite{jiang2021lisa} by invoking
SledgeHammer~\cite{sledgehammer} in each step of the proofs in the
dataset, and adding successful invocations of SledgeHammer to the
dataset. Thor trains a decoder-only transformer model (700M
parameters) on this enhanced dataset. This model is designed to learn when to invoke hammers during a proof, and guide a search process. Unlike Thor, which does not perform premise retrieval, \name{} adopts a premise retriever to enhance the performance of LLMs. Instead of performing a computationally expensive proof search, \name{} leverages LLMs to produce well-structured initial proofs and adopts repair mechanisms to fix common errors. \name{} is not directly comparable with Thor, because Thor's model is specifically trained for Isabelle proofs rather than Coq, which cannot be reproduced on Coq with reasonable effort.

Copra~\cite{thakur2023copra} uses the state-of-the-art GPT-4 model to guide a depth-first search process. In each step, GPT-4 is prompted with the proof state, previous proof steps, the incorrect steps, and the corresponding error messages to avoid recurrent errors. Copra can be further augmented by incorporating premise retrieval and generating informal proofs from informal theorem statements if they exist. However, Copra's effectiveness is highly dependent on the capability of the LLMs it uses. For example, Copra proves 26.63\% theorem in the miniF2F  dataset~\cite{zheng2021minif2f} using GPT-4, but only proves 9.02\% using GPT-3.5. Moreover, Copra does not directly repair incorrect tactics and instead prompts the LLM with incorrect tactics and error messages. In contrast, \name{} adopts a set of symbolic-based repair mechanisms to correct erroneous tactics effectively, and demonstrates consistent performance across LLMs.

\section{Conclusion}
Large Language Models (LLMs) have shown promise in automatically
generating informal proofs in natural language, but these systems have
proven to be less effective at generating formal proofs in interactive
theorem provers (ITPs). This paper described a formative study that
identified common errors made by GPT-3.5 when generating machine-checked
proofs. Guided by these findings, we proposed \name{}, which combines LLMs and symbolic methods to automatically prove theorems in an ITP. \name{} adopts a premise retriever to select relevant premises such as lemmas and definitions, in order to enhance the quality of proofs generated by an LLMs. It additionally uses a set of repair mechanisms and a backtracking algorithm to correct errors in proof scripts generated by an LLM.
% All these components are integrated within a flexible framework, that can be easily adapted to other ITPs and expanded with additional repair mechanisms.
We evaluated \name{} on a dataset of \totalnum{} theorems. In the evaluation, \name{} significantly outperforms existing approaches, and demonstrates its generalizability across different LLMs. Furthermore, our ablation study suggests that all components of \name{} are effective.
% proving between \outdsp{} and \outpassport{} more theorems. In addition, \name{} proves 1270 theorems that none of the other approaches we consider can prove.

\section*{Acknowledgements}
We thank Prasita Mukherjee and the anonymous reviewers for their valuable suggestions and feedback. This work is supported in part by NSF grants ITE-2333736, CCF-2340408, and CCF-2321680.

\bibliographystyle{ACM-Reference-Format}
\bibliography{ref}

\end{document}